\documentstyle[floats,psfig,aps,prd,amssymb]{revtex} 

\def\he#1{$^{#1}$He}
\def\li#1{$^{#1}$Li}
\def\be#1{$^{#1}$Be}

\def\h3{$^3$H}
\def\h2{$^2$H}
\def\b11{$^{11}$B}

\newcommand{\pr}{Phys. Rev. }

\newcommand{\mnras}{Mon. Not. Roy. Astron. Soc. }

\newcommand{\plb}{Phys. Lett. }
\newcommand{\Apj}{Astrophys.~J. }

\renewcommand{\phi}{\varphi}
\renewcommand{\rho}{\varrho}

\newcommand{\simge}{\gtrsim}
\newcommand{\simle}{\lesssim}

\begin{document}

\draft
\twocolumn[\hsize\textwidth\columnwidth\hsize\csname   
@twocolumnfalse\endcsname				

\title{Inhomogeneous Big Bang Nucleosynthesis: Upper Limit on $\Omega_b$
and Production of Lithium, 
Beryllium, and Boron} 

\author{Karsten Jedamzik\thanks{Electronic address: 
jedamzik@mpa-garching.mpg.de} 
and Jan B. Rehm\thanks{Electronic address: jan@mpa-garching.mpg.de} 
}
\address{
Max-Planck-Institut f\"ur Astrophysik, Karl-Schwarzschild-Str. 1,
85748 Garching, Germany
}

\date{\today}
\maketitle

\begin{abstract}
We examine the Big Bang nucleosynthesis (BBN) process in the presence of
small-scale baryon inhomogeneities. Primordial abundance yields for
D, \he4, \li6, \li7, \be9, and \b11 are computed for wide ranges
of parameters characterizing the inhomogeneities 
taking account 
of all relevant diffusive and hydrodynamic processes. 
These calculations
may be of interest due to (a) recent observations of the anisotropies
in the cosmic microwave background radiation favoring slightly larger
baryonic contribution to the critical density, $\Omega_b$, than
allowed by a standard BBN scenario and (b) new 
observational determinations of \li6 and \be9 in metal-poor halo
stars. We find considerable parameter space in which production of D
and \he4 is in agreement with observational constraints even for
$\Omega_b h^2$ a 
factor 2-3 larger than the $\Omega_b$ inferred from standard BBN. 
Nevertheless, in this parameter space synthesis of \li7 
in excess of the inferred \li7 abundance on the Spite plateau results.   
Production of \li6, \be9, and \b11 in inhomogeneous BBN scenarios
is still typically well below the abundance
of these isotopes observed in the most metal-poor stars to date thus
neither confirming nor rejecting inhomogeneous BBN. In an appendix
we summarize results of a reevaluation of baryon diffusion constants
entering inhomogeneous BBN calculations. 
\end{abstract}

]   
\section{Introduction}
The possibility that cosmic baryon number fluctuations may have
existed on small scales in the early universe  
has received considerable attention
between the late eighties and mid 
nineties~\cite{AHS:87,ibbn,JFM:94,SM:982,KKS:99}. Such fluctuations in
baryon number would have impact on the production of light
elements during BBN provided the baryonic mass of individual lumps
exceeds $M_b\simge 10^{-21}M_{\odot}\,$. It was speculated that
production of inhomogeneities could result during a first-order QCD phase
transition or even possibly during a scenario of electroweak
baryogenesis. Initially it was hoped for that such scenarios could
make BBN consistent with $\Omega_b =1$ and therefore eliminate the
need for \lq\lq exotic\rq\rq\ non-baryonic dark matter. Detailed
calculations revealed that inhomogeneous BBN (hereafter, IBBN)
scenarios may not be consistent with a universe closed in baryons due
to considerable overproduction of $^7$Li and/or \he4. At present,  
$\Omega_b =1$ seems also hardly desirable because of a variety of other
arguments, such as the cosmological baryon budget~\cite{FHP:98}, 
and the success of
a structure formation scenario employing cold dark matter, among others.  

Recently, observations of the cosmic microwave background radiation
(hereafter; CMBR) on intermediate angular scales
by the BOOMERANG and MAXIMA balloon missions have achieved
unprecedented accuracy~\cite{CMBR}. 
These missions have allowed for a first stab
at an estimate of a number of cosmological parameters such as
the total cosmic density parameter $\Omega_{tot}$
and $\Omega_b$, among others.
Common to both studies is the observation of a relatively
suppressed CMBR power spectrum on scales where a secondary peak
in the spectrum is anticipated compared to the power at the location of
the first peak. Though preliminary, the conclusion of a number of
authors is that, within the parameters commonly allowed to be
varied, an increased $\Omega_b$ could most easily
account for such a suppression~\cite{CMBOmega}. (for
alternative explanations cf. to Ref.~\cite{alter}). 
This has lead to the preliminary claim
that $\Omega_b$ as inferred from CMBR anisotropy observations
may be in conflict with the best estimate
$\Omega_b h^{2}\approx 0.02\pm 0.002$ from SBBN~\cite{BNT:00,D/H1,OSW:00}, 
in particular, the CMBR
data would prefer $\Omega_b h^2 \approx 0.03$~\cite{CMBOmega} 
($h$ is the Hubble constant in
units of 100 km s$^{-1}$Mpc$^{-1}$). Note that even though, at first glance the
deviation between these two values seems relatively small, it is
clear that a baryonic
density parameter of $\Omega_b \approx 0.03\, h^{-2}$ can 
not be achieved within a SBBN scenario. For such large $\Omega_b$, SBBN
production of deuterium can neither account for the deuterium
as observed in quasar absorption systems~\cite{D/H2,D/H1}, 
nor for the inferred D
abundance in the presolar nebula and only barely for the D 
as observed in the local interstellar medium.

Newly developed high-resolution spectrographs (such as UVES on the VLT)
allow for a significant increase in the number of stars
with claimed detections for the elements \li6 and \be9.
Whereas for a long time there had been only two claimed 
\li6/\li7~\cite{lithium6} 
detections in low-metallicity PopII halo stars, this number
is/will rapidly increase in the immediate
future. Moreover, \li6/\li7 detections
have now also been claimed for disk stars at
relatively high metallicities~\cite{Lidisk}. 
The preliminary picture which emerges is
that \li6/H abundances in stars are remarkably similar over a wide range in 
metallicities, though interpretation of the data has to account for the
possibility of stellar \li6 astration. 
Recently, there has been an interesting
\be9/H detection within the atmosphere of a very low-metallicity
star~\cite{PANH:00}. The \be9/H abundance in this star is higher than 
expected from
extrapolation of the approximately linear \be9/H versus [Fe/H]
relation such that this observation may represent 
tentative evidence for a flattening
of the \be9/H versus [Fe/H] slope at metallicities below [Fe/H] $< -3$.   

In light of the above, it seems worth reinvestigating BBN with an
inhomogeneous baryon distribution. Whereas production of \li6 and \be9
in standard BBN is essentially negligible, it is known that production of
these isotopes in IBBN may be significantly enhanced~\cite{IBBN2}. 
Furthermore,
it should be of interest
not only to find out of how much the upper limit on $\Omega_b$ in IBBN may be 
relaxed compared to that from standard BBN, but also in how much of
the parameter space, characterizing the inhomogeneities, IBBN abundance  
yields may agree with observational constraints.

\section{Inhomogeneous Big Bang Nucleosynthesis Calculations} 
We have performed detailed numerical computations of IBBN by employing
the IBBN code described in Ref~\cite{JFM:94}. 
This code treats all the relevant baryon
diffusion of neutrons, protons, and lighter nuclei. In the Appendix we
summarize the employed baryon diffusion constants for protons and
neutrons, which includes a reevaluation of some diffusion constants and 
a correction for mistakes in the literature. The employed code is 
still the only existing code with a detailed treatment of the effects
of photon diffusion and hydrodynamic expansion on the evolution of 
high-density regions~\cite{ADFMM:90,JF:94}. 
It is known that these dissipative
processes operating at lower temperatures $T\simle 30\,$keV
may affect the predicted
abundances of \li6, \li7, \be9, and \b11 in some part of the parameter
space (in particular, for compact high-density regions).
For example, \li7 produced in form of \be7 may be prematurely
destroyed by the reaction sequence \be7 $(n,p)$\li7 $(p,\alpha
)\alpha$ when enough neutrons may be delivered to the high-density
regions where most of the \be7 is produced. The magnitude of this process
depends on the efficiency of hydrodynamic expansion which increases the
surface area of high-density regions but also on the correct neutron-
and proton- diffusion constants at low temperatures. Note that the
distribution (and diffusion) of protons affects the diffusion of
neutrons through neutron-proton nuclear scattering~\cite{remark2}.

We have updated the nuclear reaction rates employed in the IBBN code 
from those based on the compilation by Caughlan \& Fowler, as
described in Smith {\it et al.}~\cite{SKM:93}, 
to include the improved charged nuclei induced
reactions as compiled by the NACRE~collaboration \cite{NACRE}. 
Note that the modifications in predicted
abundances, when the central values of the
improved nuclear reaction rates of the NACRE
compilation are employed, are fairly small for \h2 and \he4 but
can be in the $\sim$ 20\% - 30\%  range for \li7, \li6, \be9, and
\b11 (cf. to Ref.~\cite{VCC:00}). 
Additional uncertainties in the computed abundances arise from 
appreciable error bars quoted in the NACRE compilations for a few
reactions, such as D(p,$\gamma$)$^3$He, $^3$He($\alpha ,\gamma$)$^7$Be,
and D($\alpha ,\gamma$)$^6$Li. In the context of SBBN, these 
additional uncertainties are of
similar magnitude to those quoted above, with the exception of the
$^6$Li abundance which is subject to very large uncertainties
of $\sim$ factor 3-4 in either direction~\cite{NLS:97,VCC:00}.
Though it is beyond the scope of the present work to present a detailed
systematic analysis of uncertainties in the prediction of abundances
in IBBN scenarios due to reaction rate uncertainties, 
we will comment below if such
uncertainties could impact our main conclusions. 
Note that all \he4 mass fractions $Y_p$  presented below 
are corrected by $\Delta Y_p = +0.0049$ to account for a variety of 
physical effects as detailed in Ref.~\cite{LT:99}.

From the multitude of conceivable initial conditions for the
baryon inhomogeneities (including stochastic ones as treated in 
Ref.~\cite{KJM:97}) 
we chose a regular lattice of spherical symmetric domains, approximating
the possible outcome of baryon fluctuations generated during a
first-order (e.g. QCD) phase transition around the shrinking bubbles of
high-temperature phase. The spherical computation domain is then 
characterized by its physical length, $l_{100}$, specified at
temperature of $T=100\,$MeV (specifically, 1 m at $T=100\,$MeV  
is to be understood as a length of $5.96 \times 10^{11}$m
at the present epoch~\cite{remark3}).
Within this domain we assume a region of high baryon density with 
baryon-to-photon ratio $\eta_h$ occupying volume fraction $f_V$ 
and a low density region at $\eta_l = \eta_h/R$ occupying the remainder of the
volume, with an initial discontinuity at the boundary of 
both regions (which softens after some baryon diffusion).
This yields an average baryon-to-photon ratio
\begin{equation}
\eta = f_V R\,\eta_l + (1-f_V)\,\eta_l\, .  
\end{equation}
Given these initial conditions there exist still four parameters to be
specified, namely $\eta$, $l_{100}$, $f_V$, and $R$. The initial
parameter space is reduced by assuming $f_V R = 200$. Physically 
$f_V R \gg 1$ corresponds to essentially all baryons residing in
the high-density region and none in the low-density region, such that
for $f_V R \simge 10$ one obtains results essentially
independent of the exact value of this parameter combination. Though
the opposite limit, $f_V R \ll 1$, may be interesting for the
production of significant amounts of isotopes with nucleon number
$A\geq 12$~\cite{JFMK:94}, in much of the parameter space it yields only minor
changes in the D, $^4$He, and $^7$Li as compared to a SBBN scenario
at the same $\eta$. Our calculations employ two different initial 
\lq\lq geometries\rq\rq : (a) spherical condensed - where the
high-density region resides at the center of the spherical domain 
and (b) spherical shell - where the high-density region occupies a 
shell at the outer edge of the computational domain. These spherical 
domains are finite-differenced into 24 zones. By increasing the number
of zones
we estimate that the relative error in predicted abundances 
does not exceed $\sim 0.5$\% for $^4$He, $\sim 3$\% for D, 
and $\sim$ 10\% for the shown isotopes with $A\geq 7$.

\section{Results and Discussion}

Figures 1 - 4 show computed abundance yields in IBBN scenarios for
the isotopes of D, \he4, \li7, as well as \li6, \be9, and \b11 as
a function of the length scale of the domains (approximately
corresponding to the mean separation between fluctuations) for 
differing $\Omega_b$ and a wide range of parameters describing the 
baryon inhomogeneities. The choice of the parameter space for which 
abundance yields are shown is supposed 
to bracket most potentially interesting $\Omega_b$ 
(taking values of $\Omega_b h^2= 0.012$, $0.025$, $0.038$ and $0.051$ 
shown by the solid, dotted,
short-dashed, and long-dashed lines in each figure, respectively), 
as well as to illustrate the general trends of changing 
\lq\lq geometry\rq\rq\ (Figures 1 and 2 are for spherical condensed 
fluctuations, whereas Figure 3 and 4 are for spherical shells)
and changing volume fractions $f_V$ of the high-density regions.
We have deliberately not indicated observationally inferred limits on
the primordial abundances in these figures, 
as these are likely to change over the
course of time, and since we are more interested in a qualitative
understanding of IBBN abundance yields and their potential 
agreement/disagreement with observationally inferred abundance limits.

 \begin{figure*}					
    \centerline{\psfig{figure=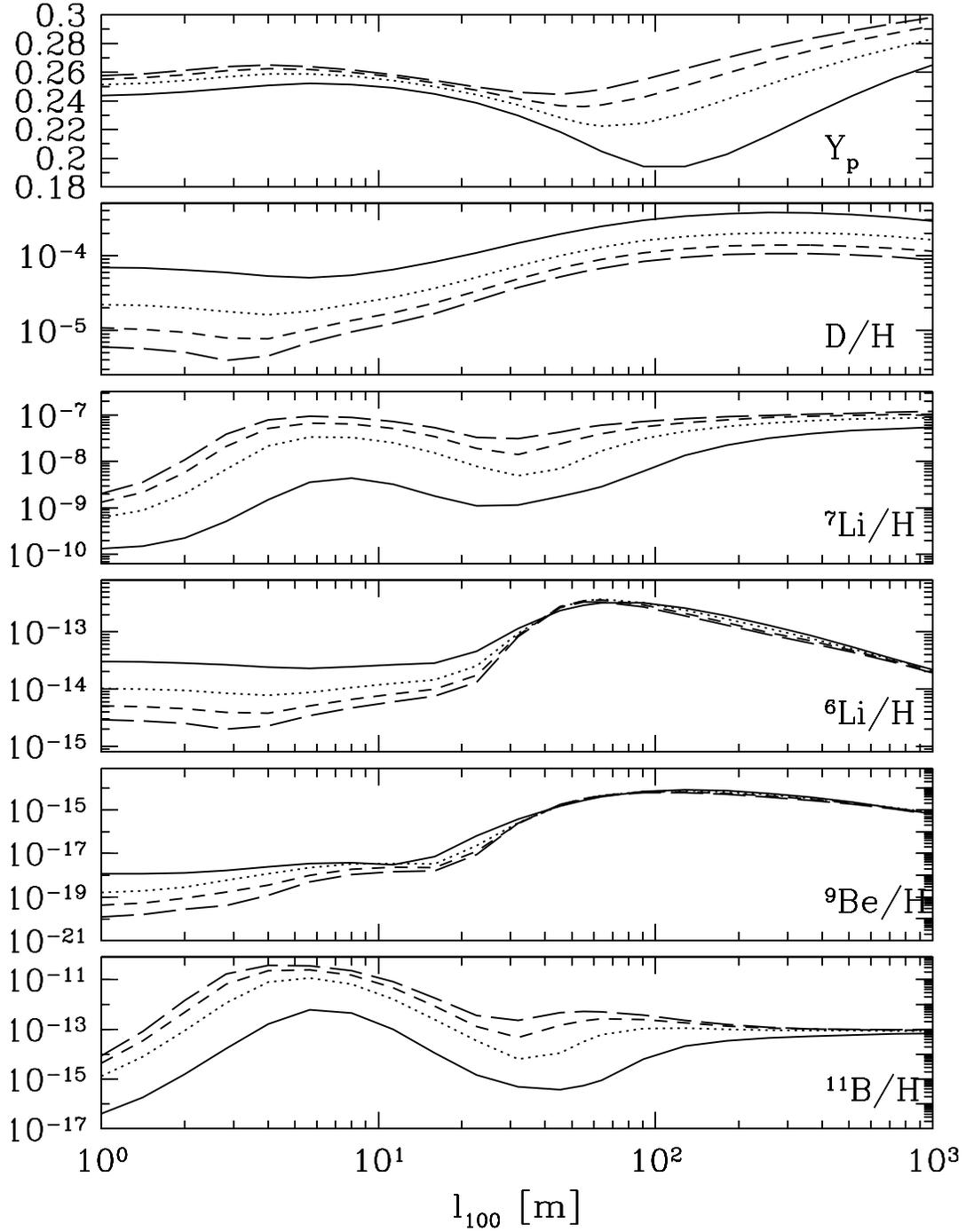,height=20cm}}	
 {\caption{Abundance yields of D, \he4, \li6, \li7, \be9, and \b11  
in IBBN scenarios as a function of the inhomogeneity length scale  
$l_{100}$ (given in meters at $T=100\,$MeV). The calculation  
assumes spherical condensed  
inhomogeneities with high-density volume filling fraction $f_V = 
0.125^3$ and density contrast between high- and low- 
density regions of $R = 200/f_V$ (see text for details).  
Except for the \he4 abundance which is given as mass fraction, $Y_p$, 
all abundances are given as number fractions relative to 
hydrogen as indicated in the panels.  
The solid, dotted, short-dashed, and long-dashed lines refer 
to results for $\Omega_b h^2 = 0.012$, $0.025$, $0.038$, and 
$0.051$, respectively.  
   \label{F:core_125}}} 
 \end{figure*} 

 \begin{figure*}  
    \centerline{\psfig{figure=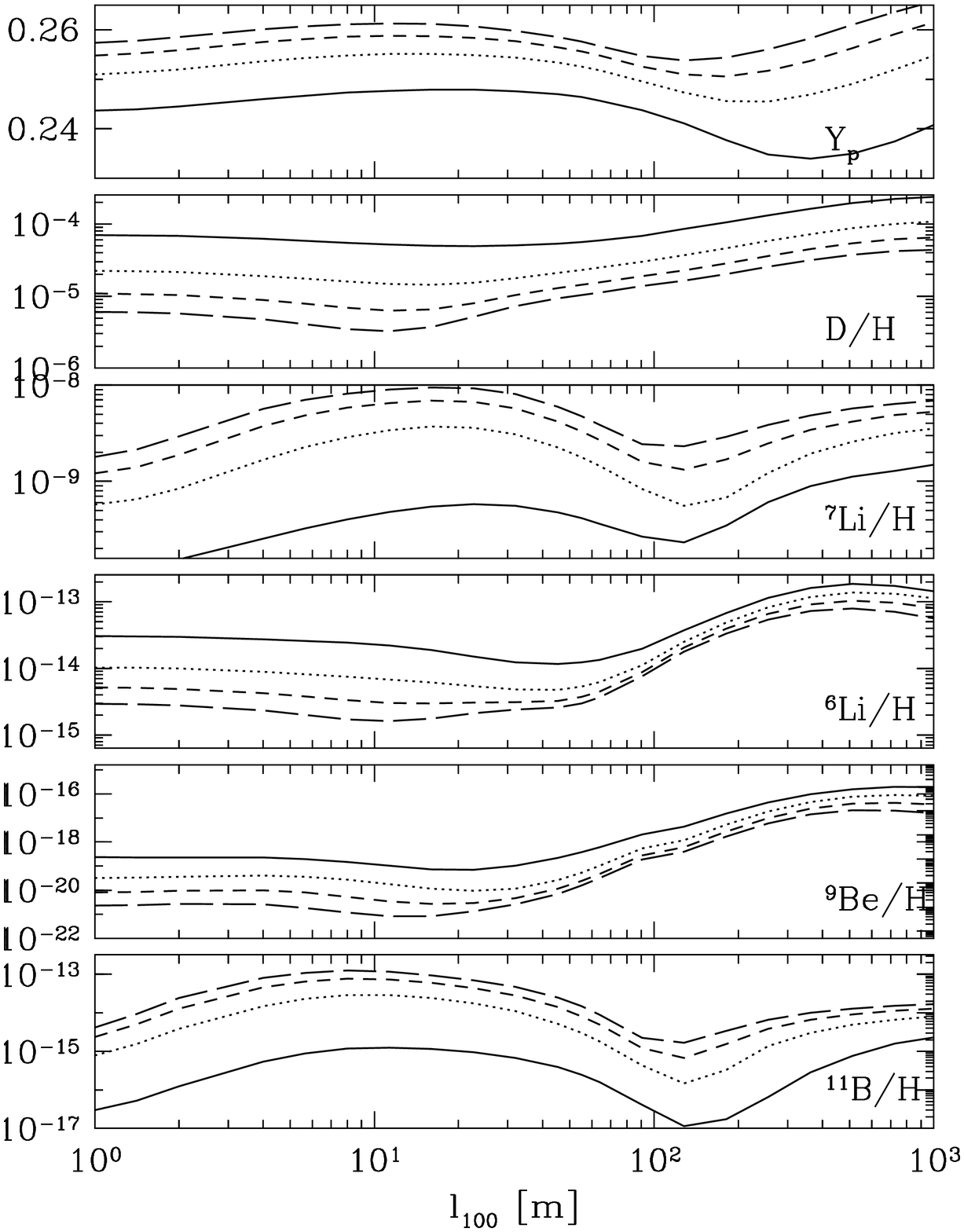,height=20cm}} 
 {\caption{As Figure 1, but for $f_V = 0.5^3$.   
   \label{F:core_500}}} 
 \end{figure*} 

It is well known that there exists an \lq\lq optimum\rq\rq\ length
scale where the \he4 yield may be lower than in a SBBN scenario at the
same $\Omega_b$ due to
the fact that neutrons may diffuse out of the high-density regions,
subsequently decaying in the low-density region
before they may be 
incorporated into \he4. This effect is more pronounced
when $f_V$ is small (Figures 1 and 3)  since back-diffusion of 
neutrons into the high-density regions is less efficient.
Similarly, one finds an \lq\lq optimum\rq\rq\ $l_{100}$ where \li7
production is minimized, generally somewhat smaller than the length
scale for minimum \he4 production. Nevertheless, even at this
\lq\lq optimum\rq\rq\ distance the \li7 yields are often higher, at best
somewhat lower, than the \li7 yields in a SBBN scenario at the
same $\Omega_b$ and they increase with decreasing 
$f_V$~\cite{remark2}. These trends
are due to the two different production mechanisms for \li7 (direct
production in the low-density region and production of \be7  
in the high-density region) and the relative efficiency of these mechanisms
at either lower or higher $\eta$ than the approximate $\eta$ 
inferred from SBBN. 

There are currently two mutually inconsistent observationally inferred
values for the \he4 mass fraction, i.e. a high value 
$Y_p \approx 0.244$~\cite{IT:98} and a low value 
$Y_p \approx 0.234$~\cite{OS:97,PPR:00}. 
It also becomes more and more appreciated that
the inference of $Y_p$ from observations of HII regions is subject to
systematic errors of possibly considerable magnitude. 
Much progress has been made in the determination of primordial D
abundances in quasar absorption line systems (QASs). There are now several
QASs seemingly indicating low D/H 
$\approx 2.5 - 4\times 10^{-5}$~\cite{D/H2,D/H1} and
only one which favors high D/H 
$\approx 2\times 10^{-4}$~\cite{We:97}. In light of
this, one should probably demand from a sucessful BBN scenario to have
$Y_p < 0.25$ (conservative) and low D/H $\approx 2 - 5\times 10^{-5}$.
(Note that even within the context of a SBBN scenario, a D/H $\approx
3\times  10^{-5}$ abundance implies seemingly uncomfortably
high $Y_p\approx 0.247$~\cite{BNT:00,D/H1}.
The figures illustrate that 
one may find considerable IBBN parameter space where these
requirements on the primordial \he4 and D/H abundances may be met, even
for $\Omega_b h^2$ as large as $\sim 0.05$. This is due to IBBN scenarios
often yielding  
less \he4 and more D production than a SBBN scenario
at the same $\Omega_b$.

 \begin{figure*} 
    \centerline{\psfig{figure=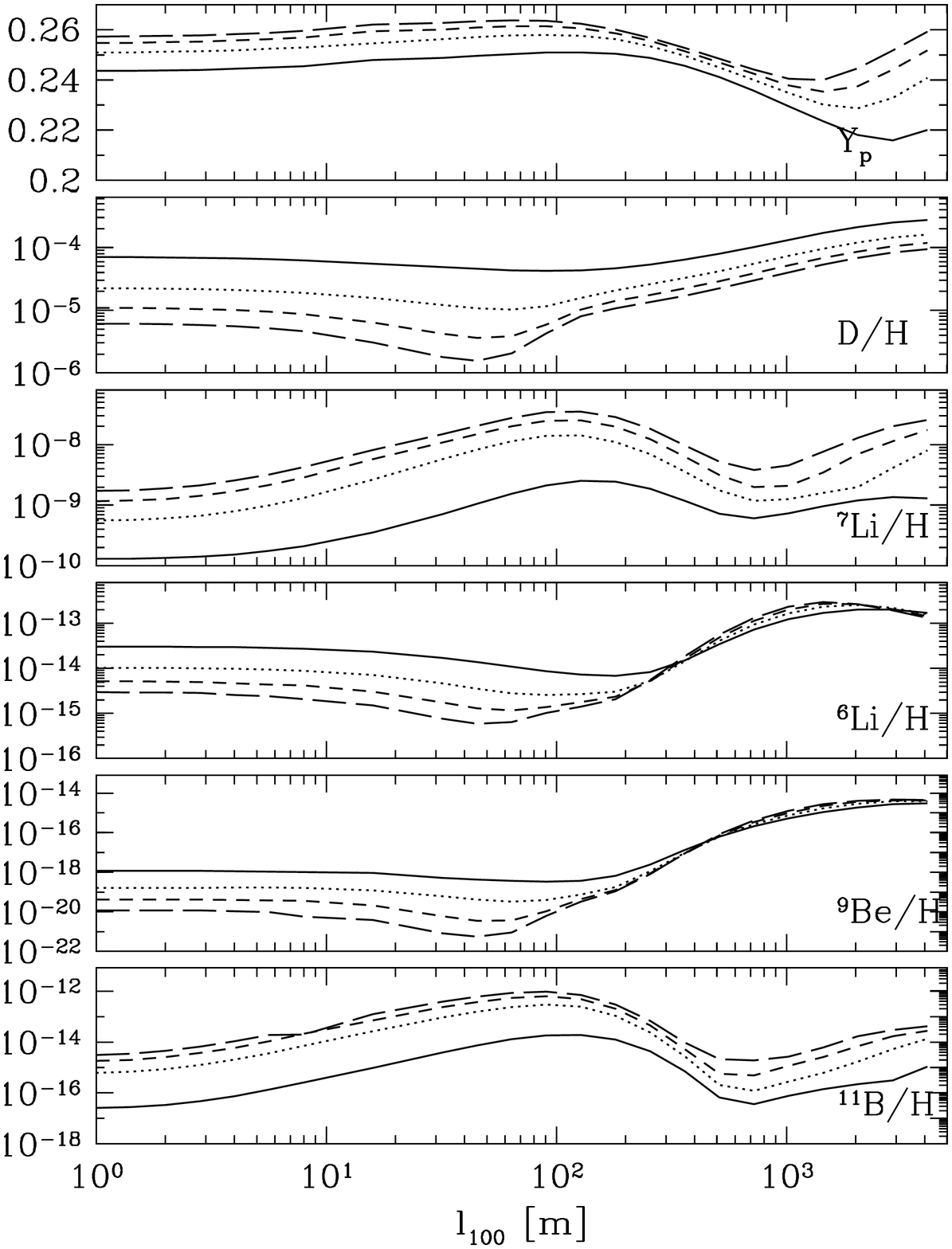,height=20cm}} 
 {\caption{As Figure 1, but for spherical shell geometry and $f_V =  
    0.25^3$.  
   \label{F:shell_250}}} 
 \end{figure*} 

 \begin{figure*}  
    \centerline{\psfig{figure=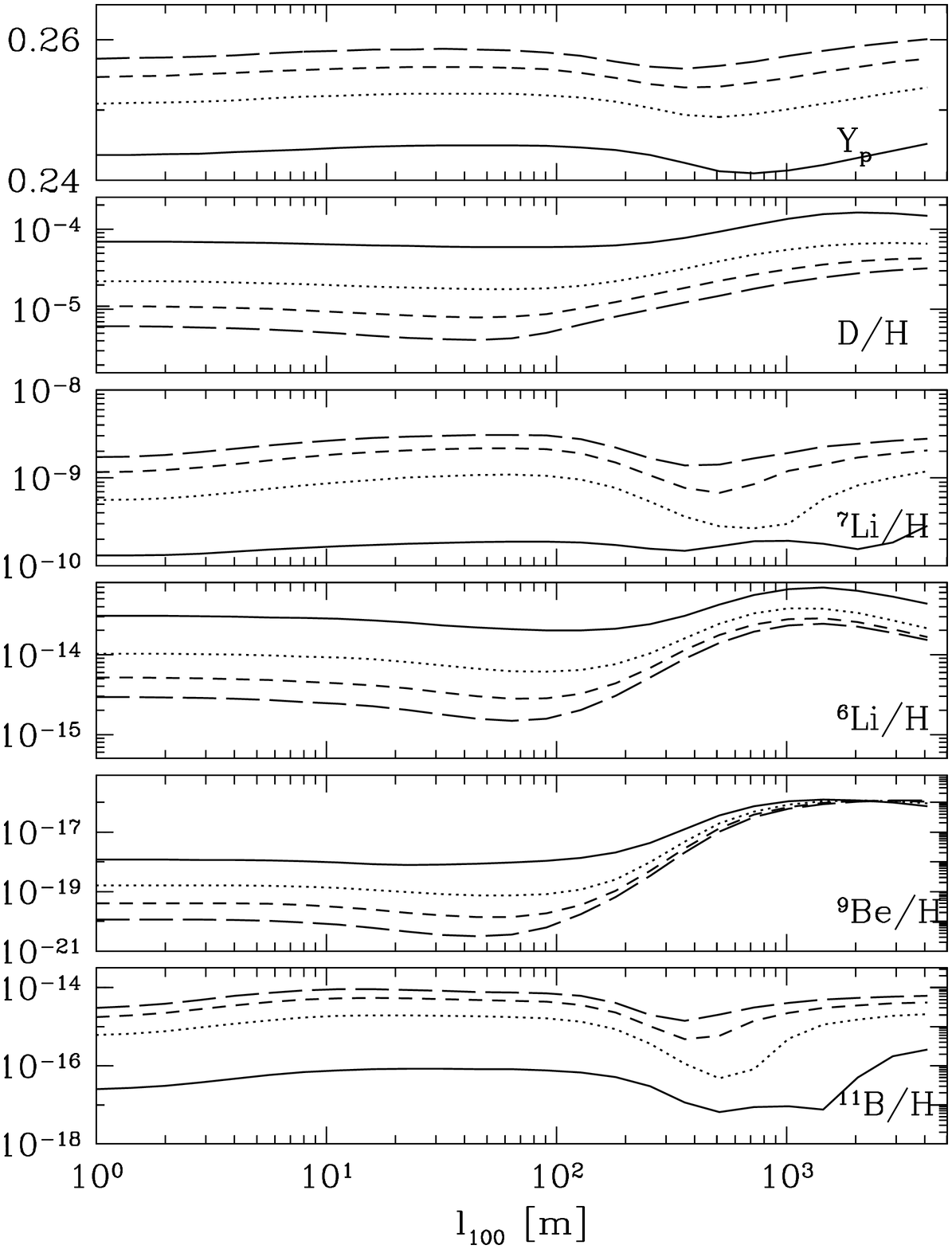,height=20cm}} 
 {\caption{As Figure 1, but for spherical shell geometry and $f_V =  
    0.8^3$.   
   \label{F:shell_800}}}  
 \end{figure*} 

Nevertheless, these considerations disregard observational limits
on the \li7/H abundance.
Typical \li7 yields in the IBBN parameter space which agree with
observational limits on $Y_p$ and D/H strongly depend on $\Omega_b$
(as well as on geometry and $f_V$), ranging between about
$3\times 10^{-10}$ and $10^{-8}$ for $\Omega_b h^2 = 0.025$ and
$3\times 10^{-9}$ to $3\times 10^{-8}$ for $\Omega_b h^2 = 0.051$,
This is typically well in excess of the claimed primordial
$^7$Li/H $\approx 1.7\times 10^{-10}$ as derived from observations
of lithium abundances in metal-poor halo stars belonging to the Spite 
plateau. Though \li7 may in principle be depleted in these stars, there
are strong arguments against this possibility, such as the claimed
absence of intrinsic dispersion of \li7 abundances in stars belonging
to the Spite plateau. Further, the more fragile \li6 isotope which should have
been astrated  as well, is by now observed in  a few of these stars
\cite{LSTC:97}. Recently,  
a primordial \li7/H even as low as $1.2\times 10^{-10}$ 
has been claimed~\cite{Retal:00}, which results from
correcting \li7 abundances for galactic cosmic ray production of this
isotope. Nevertheless, even an SBBN scenario (at $\Omega_b h^2 =0.02$)
yields \li7/H 
$\approx 3.8\times 10^{-10}$, in excess of the primordial
\li7/H determination, when one 
assumes D/H $3\times 10^{-5}$ 
as favored by the QAS data.
One thus would have to resort to a small amount
of \li7 depletion and/or systematic errors due to, for example, the use
of inappropriate stellar atmospheric models, to reconcile these values.
If one demands the IBBN yield of \li7 to be at, or below, the quoted
SBBN reference value, one finds that the $\Omega_b h^2$ should be
below $0.025$, precluding a substantial increase of $\Omega_b$ over
that inferred from SBBN. Only if one were to relax the 
\li7/H limit to about (seemingly unreasonable) $10^{-9}$ could 
$\Omega_b h^2$ in IBBN scenarios
be consistent with the currently preferred $\Omega_b h^2$
from CMBR anisotropy measurements $\sim 0.03$.

We have tested if these conclusions could be changed
due to existing reaction rate uncertainties quoted by
the NACRE collaboration~\cite{NACRE}. We have changed the following
reactions to their quoted limits: D(p,$\gamma$)\he3 (lower limit),
$^3$H($\alpha ,\gamma$)$^7$Li (lower limit), \he3($\alpha
,\gamma$)$^7$Be (lower limit), \li7(p,$\alpha$)\he4 (upper limit), and
D($\alpha ,\gamma$)$^6$Li (upper limit). These changes have been
\lq\lq designed\rq\rq\ to minimize \li7 production and maximize \li6
production. With these modified rates we have performed two
calculations (a) spherical shell, $\Omega_b h^2=0.038$, $f_V^{1/3}=0.025$,
$l_{100}=724\,$m, and (b) spherical condensed, $\Omega_b h^2=0.038$, 
$f_V^{1/3}=0.0125$, $l_{100}=32\,$m, where the length scales have been
chosen close to the \lq\lq optimum\rq\rq\ distance for minimum \li7 
production. This has lead to a \li7/H yield of 
$1.45\times 10^{-9}$ (compared to $2\times 10^{-9}$ when the central
values of the reaction rates are used) in case (a)
and $1.02\times 10^{-8}$ (compared to $1.4\times 10^{-8}$) in case
(b), illustrating that the uncertainty in the predicted
\li7 remains within bounds. 
Nevertheless, a large uncertainty exists in the prediction for the
\li6 abundance: we have found a factor $\sim 3$ and 4 increase in case
(a) and (b), respectively.  

It should be interesting to explore if the abundance yields of \li6,
\be9, and \b11 in IBBN scenarios may be as large as the abundances of
these isotopes observed in the most metal-poor stars to date.   
Such a comparison could yield, in principle, independent 
confirmation/rejection of IBBN scenarios. In the parameter space where
IBBN yields are consistent with observationally inferred limits on
\he4 and D/H, we find production of \li6/H $\sim 10^{-14} -
7\times 10^{-13}$ (including the large reaction rate uncertainty
in the D($\alpha ,\gamma )$\li6 rate), 
implying a typical \lq\lq maximum\rq\rq\
enhancement factor for this isotope
of about 10-30 compared to a SBBN scenario at the same $\Omega_b$.
For the \be9 and \b11 isotopes one finds ranges of 
\be9/H $\sim 10^{-18} - {\rm a\, few\,}\times 10^{-15}$, and
\b11/H $\sim 10^{-16} - 10^{-13}$. Typical IBBN yields of these
isotopes seem therefore still much below the observed \li6/H $\sim 7\times
10^{-12}$~\cite{lithium6}, \be9/H 
$\sim 5\times 10^{-14}$~\cite{Betal:99,PANH:00}, and 
\b11/H $\sim 10^{-12}$~\cite{GLetal:98}
in the lowest metallicity stars 
to date where such observations have been performed.
These observations are thus inconclusive with regards to a validation
of IBBN scenarios.
Though it is not easy to completely rule out the possibility that there 
indeed exist very specific initial conditions for the baryon inhomogeneities
which yield primordial production of \li6, \be9, and \b11 in
abundance as high as currently observed in the lowest metallicity stars, it
seems clear that this is not the typical case.

In summary, we have performed numerical simulations of 
BBN in the presence of an inhomogeneous baryon distribution for
wide ranges of the parameters describing the inhomogeneities and for
a few representive baryon-to-photon ratios. Our choice of initial
conditions is limited to scenarios where essentially all baryons are
within overdense pockets and the remainder of the volume is initially
void of baryons. We found that such scenarios may be consistent with
observational limits on the primordial \he4 and D abundance 
for $\Omega_b h^2$ as large as $\sim 0.05$, however, they
result in significantly 
overabundant production of \li7 with respect to the \li7/H
ratio as observed in stars belonging to the Spite plateau.
We note here that similar conclusions have been recently 
drawn by Ref.~\cite{KS:99}. 
Typical production of \li6, \be9, and \b11 in such scenarios are found
to be still below the abundances of these isotopes observed in the
most metal-poor stars to date. Unless \li7 in stars on the Spite
plateau has been significantly astrated, which seems unlikely,
IBBN scenarios thus do not allow for a significant increase of $\Omega_b$
over that inferred from a SBBN scenario.

\vskip 0.15in
We wish to acknowledge several useful discussions with
In-Saeng Suh and Naoki Yoshida.

\appendix

\section{Reevaluation of Baryon Diffusion Constants}\label{S:Appendix}

In this appendix we summarize the baryon diffusion constants
which we used in our inhomogeneous Big Bang nucleosynthesis
calculations. Some of these diffusion constants have been reevaluated.
Such a reevaluation seemed necessary not only since
prior work on the subject~\cite{AHS:87,BC:91} yielded partially 
conflicting results
(e.g., the proton diffusion constant due to proton-electron scattering
as computed by Applegate, Hogan, \& Scherrer and Banerjee \& Chitre),
but also due to improvement on approximations, such as an energy
independent neutron-proton cross section. Furthermore, we correct for
the electron diffusion constant due to electron-photon scattering as
given in~\cite{JF:94}.
Rather than going over the partially lengthy details of the
calculations we performed, we will state our results, outline by what 
procedure we obtained them, and highlight the differences to prior
evaluations.

Banerjee \& Chitre~\cite{BC:91} 
(hereafter; BC) computed diffusion constants by using the
first-order Chapman-Enskog approximation for arbitrarily relativistic
particles as thoroughly discussed in the monograph by de Groot, van
Leeuwen, \& van Weert~\cite{GLW} (hereafter; GLW). The master equation
given in BC for the computation of diffusion constants (i.e., Eqs.~(1),
(3), and (4) in BC) are not directly evident from GLW but involve a fairly
detailed computation. We have therefore redone the calculation of
this master equation and arrive at the same
result~\cite{remark1} as BC. Note
that the first-order Chapman-Enskog approximation is typically
accurate to within 20-30 \%~\cite{GLW}. 

\vskip 0.1in
\noindent
{\bf (a) neutron-electron scattering}
\vskip 0.07in

At higher temperatures ($T \simge 50 - 100\,$keV), and
when the local baryon-to-photon ratio ($\eta$) is not too large,
the diffusion of neutrons is limited by magnetic moment scattering off
electrons and positrons. Using the master equations of BC, under the
assumption of an energy-independent cross section, and to lowest
non-trivial
order in the small quantities $m_e/m_N$ and $T/m_N$, where $m_e$,
$m_N$, and $T$ are electron mass, nucleon mass, and temperature,
respectively, but for arbitrary $T/m_e$, we find
\begin{equation}
D_{ne} = {3\over 8}\sqrt{{\pi\over 2}}{1\over
n_{e^{\pm}}\sigma_{ne}^t}{1\over
z_e^{1/2}}{K_2(z_e)\over K_{5/2}(z_e)}\, ,\label{E:Dne}
\end{equation}
in agreement with BC. In this expression $z_e = m_e/T$, the quantity
$n_{e^{\pm}}$
is the total number density of electrons and positrons,
the transport cross section is
\begin{equation}
\sigma_{ne}^t = \int d\Omega {d\sigma_{ne}\over d\Omega} (1-cos\theta )\, ,
\end{equation}  
to be evaluated in the center-of-mass system,
and $K_n$ are modified Bessel functions of the second kind and of n'th
order, i.e.
\begin{equation}
K_2(z) = {1\over z^2}\int_0^{\infty} k^2\, {\rm exp}\,
({-\sqrt{k^2 + z^2}})\,{\rm d}k\, ,
\end{equation}
and
\begin{equation}
K_{5/2}(z) = \sqrt{\pi\over 2z}e^{-z}\biggl(1 +{3\over z} + {3\over
z^2}\biggr)\, .
\end{equation}
The expression Eq.~(\ref{E:Dne}) does agree with that derived by 
Applegate, Hogan,
\& Scherrer~\cite{AHS:87} (hereafter; AHS) 
via considering the drag force exerted by
$e^{\pm}$ on neutrons and using the Einstein relation. 
As noted by Ref.~\cite{KAGMBC:92}, 
both derivations of the diffusion constant approximate the fermionic
occupation number by a relativistic Maxwellian, i.e. 
$f = ({\rm exp}(E/T) + 1)^{-1}\approx {\rm exp}(-E/T)$, 
which nevertheless should
only result in a small error.
Using $\sigma_{ne}^t = 3\pi\, (\alpha\kappa /m_N)^2$, with
$\alpha$ the fine structure constant and $\kappa = -1.91$, we may give 
a numerical value for the diffusion constant 
\begin{equation}
D_{ne} = 1.87\times 10^4 {\rm {m^2\over s}}\biggl({T\over {\rm
MeV}}\biggr)^{1/2} {1\over (n_{e^{\pm}}/{\rm MeV^3})}{K_2(z_e)\over
K_{5/2}(z_e)}\, ,
\end{equation}
where $n_{e^{\pm}}$ is given in natural units, 
i.e. $\hbar = c = 1$. The correct density $n_{e^{\pm}}$ is easily
obtained from the BBN code.

Recently Suh \& Mathews~\cite{SM:981,SM:982} have considered finite-temperature
effects on the neutron diffusion constant. They find a transport cross
section due to neutron-electron (positron) magnetic moment scattering which
significantly increases over $\sigma_{ne}^t = 3\pi (\alpha\kappa /m_N)^2$
at low $T \simle 0.5\,$MeV. Since finite-temperature effects should
vanish in the limit $T\to 0$ their result is fairly surprising. We have
therefore reevaluated the neutron-electron cross section by using the
result given in Ref.~\cite{Pil} and confirm that neutron-electron
scattering
is independent of energy for electron energies much below the nucleon
mass. Similarly, the authors claim a significant increase of $D_{ne}$
at high $T$~\cite{SM:982}. Within the context of their analysis, such an
increase could only occur due to a change in the electron mass and/or the
transport cross section. Nevertheless, according to their own analysis
both quantities don't seem to deviate much from their zero-temperature
limits for temperatures below $T\simle 3-5\,$MeV. In light of these
inconsistencies we therefore prefer 
to use the standard AHS and BC results in our calculations.

\vskip 0.1in
\noindent
{\bf (b) neutron-proton scattering}
\vskip 0.07in

Neutron-proton nuclear scattering limits diffusion of neutrons at
lower temperatures and/or high $\eta$. At the low energies relevant
for BBN the scattering cross section is dominated by scattering-angle
independent s-wave scattering
(zero angular momentum) resulting in a transport cross section which
equals the total cross section,
\begin{equation}
\sigma_{np} = {\pi a_s^2\over (a_s k)^2 + (1 - {1\over 2}r_s a_s
k^2)^2} + {3\pi a_t^2\over (a_t k)^2 + (1 - {1\over 2}r_t a_t
k^2)^2}\, .\label{E:snp}
\end{equation}
In this expression $k$ is the nucleon wave vector in the
center-of-mass (!) system~\cite{BW}, and the parameters $a_s =
-23.71\,$fm, $a_t = 5.432\,$fm, $r_s = 2.73\,$fm, and $r_t=1.749\,$fm.
Only at very low temperatures the above cross section is approximately
independent of energy. In that case, one may derive~\cite{GLW}
\begin{equation}
D_{np} = {3\sqrt{\pi}\over 8}{1\over n_p\sigma_{np}}
\biggl({T\over m_N}\biggr)^{1/2}\, ,\label{E:dnp}
\end{equation}
for the diffusion constant, which is a factor of two smaller 
than the result given in BC.
Nevertheless, the energy dependence of the cross section becomes large
at $T\simge 50\,$keV and should be taken into account. This may be
done properly by evaluating the diffusion constant via the master 
equations in BC with the appropriate cross section Eq.(\ref{E:snp}). 
Following this procedure we derived the cross section
\begin{eqnarray}
D_{np} = {\rm 2.82\times 10^{-5}{m^2\over s}}\biggl({T\over {\rm
MeV}}\biggr)^{1/2}{1\over (n_p/{\rm MeV^3})}\times \nonumber\\
\times {1\over I(a_1,b_1) + 0.16
I(a_2,b_2)}\quad\quad\quad\, ,\label{E:dnpcorr}
\end{eqnarray}
where
\begin{equation}
I(a,b) = {1\over 2}\int_0^{\infty}{\rm d}x {x^2 e^{-x}\over a x + (1
- b x/2)^2}\, ,
\end{equation}
are integrals to be evaluated numerically. Here the parameters $a$ and
$b$ are given by
$a_1 = 13.59\, ({T/ {\rm MeV}})$, $b_1 = -1.56\, ({T/ {\rm
MeV}})$, $a_2 = 0.71\, ({T/ {\rm MeV}})$, and $b_2 = 0.23\, ({T/ {\rm
MeV}})$. We note that the integral $I(a,b)$ converges only slowly
against its limiting value $I(0,0) = 1$ as the temperature is decreased.
In that limit, Eq.~(\ref{E:dnpcorr}) converges against Eq.~(\ref{E:dnp})
with $\sigma_{np} = \pi a_s^2 + 3\pi a_t^2$.
\vskip 0.1in
\noindent
{\bf (c) proton-electron scattering}
\vskip 0.07in
The diffusion of protons in IBBN scenarios is only significant
at lower temperatures. As long as Debeye screening of proton charge in the
plasma is effective, protons may diffuse independently of the additional
(net) electrons required by charge neutrality 
(cf. electron-photon scattering).
In that case, proton diffusion is limited by Coulomb scattering on
$e^{\pm}$.
Both, AHS and BC compute the proton diffusion constant due to Coulomb
scattering $D_{pe}$. However, their results differ by as
much as a factor of eight at low temperatures. We have therefore
recomputed $D_{pe}$ by using the master equations given in BC and
employing the Mott scattering cross section. To lowest order in
$m_e/m_N$, and accurate to first order in $T/m_e$, we obtain
\begin{equation}
D_{pe} = {3\over 4\sqrt{2\pi}}{T^2\over
\alpha^2n_{e^{\pm}}}\biggl({T\over
m_e}\biggr)^{1/2}{1+{15\over 8}{T\over m_e}\over 
\Lambda + 2{T\over m_e}(\Lambda
-1)}\, ,\label{E:dpe}
\end{equation}
where $\Lambda$ is the well-known Coulomb logarithm with
$\Lambda \approx {\rm ln}\, (T^2m_e/2\pi\alpha n_{e^{\pm}})^{1/2}$.
In the limit $T/m_e\to 0$ Eq.~(\ref{E:dpe}) reproduces the result of
AHS. We conclude that $D_{pe}$ as calculated by BC is a factor of eight
to small at low $T$. Deviations between Eq.~(\ref{E:dpe}) and the result
of AHS to first order in $T/m_e$ are due to AHS approximating the
electron energy in 
the Mott scattering cross section by $m_e$. 
Eq.~(\ref{E:dpe}) yields for the numerical value of 
the proton diffusion constant
\begin{eqnarray}
D_{pe} = 9.29\times 10^{-2}{\rm {m^2\over s}}
\biggl({T\over {\rm MeV}}\biggr)^{5/2}{1\over (n_{e^{\pm}}/{\rm
MeV^3})}\times\nonumber\\ \times {1 + {15\over 8}{1\over z_e}\over
{(\Lambda /5)+ {2\over z_e}\, ((\Lambda /5)- 1/5)}}\, .
\end{eqnarray}
\vskip 0.1in
\noindent
{\bf (d) electron-photon scattering}
\vskip 0.07in
With the decrease of temperature thermally produced electron-positron
pairs become rare and Debeye screening of nuclear charge becomes
inefficient. In this limit, electric forces which would rapidly be
build up if the proton- and (net) electron- distributions differed,
prevent the independent diffusion of these two species~\cite{JF:94}. One
may show, by evaluating the electric fields which would be present
due to differing proton- and electron- distributions in the presence of
Debeye screening, and by comparison of the resulting proton flux due
to the electric fields with the flux of protons due to diffusion, that
protons may only diffuse independently, when $n_{e^{\pm}} \gg n_{e^-}
- n_{e^+}$. 
When this is not the case, electrons
and protons diffuse together  by ambipolar diffusion, with the effective
diffusion constant given by twice that of the larger of electron- and
proton- diffusion constants~\cite{LL}.
Electron diffusion is rendered fairly inefficient due to Thomson
scattering of electrons on the cosmic background photons. The
diffusion constant may be computed by considering the drag force on
an electron due to a photon blackbody~\cite{Peeb}
\begin{equation}
f_{\rm drag} = {4\over 3}{\pi^2\over 15}\sigma_{\rm Th} T^4 v\, ,\label{E:fdrag}
\end{equation}
with $\sigma_{\rm Th} \approx 6.65\times 10^{-29}\,{\rm m^2}$ the Thomson cross
section and $v$ the velocity of the electron in the cosmic background
photon rest frame. Note that Eq.~(\ref{E:fdrag}) is given in natural
units. Using Eq.~(\ref{E:fdrag}) together with the Einstein relation
$D = T b$, where the mobility $b$ is defined as the proportionality
constant between the terminal velocity $v$ which a particle reaches 
in a plasma when an external force $f$ is applied, i.e. $v = b f$,
one finds for the effective proton diffusion constant
\begin{equation}
D_p^{\rm eff} = 2 D_e = 7.86\times 10^{-2}{\rm {m^2\over s}}
\biggl({T\over {\rm MeV}}\biggr)^{-3}\, ,\label{E:dpeff}
\end{equation}
applicable when $n_{e^+} < n_{e^-}$. We note here that
Eq.~(\ref{E:dpeff}) is different from the simple estimate given in 
Ref.~\cite{JF:94}.
We stress that the neglect of $D_p^{\rm eff}$ due to Thomson scattering,
typically important at low $T\simle 40\,$keV,
may lead to errors in the calculated $^7$Li abundances by more than
an order of magnitude.

\newcommand{\singleletter}[1]{#1}






\begin{thebibliography}{10}
\bibitem{remark3} Here we assumed a statistical weight of $g_{\rm eff} =
10.75$ at $T=100\,$MeV and $g_{\rm eff} = 3.909$ at the present epoch, 
thus neglected muon and pion annihilation
and assumed the completion of the cosmic quark-gluon transition.
\bibitem{remark2} We have found that production of \be7 may be
reduced by more than an order of magnitude
if one (incorrectly) neglects the contribution of Thomson scattering
to the proton diffusion constant (cf. Appendix). This may explain
the discrepancies between our conclusions to be drawn below and the
conclusions reached in Ref.~\cite{SM:982}.
\bibitem{remark1}
Note that in Eq.(4) of BC $d\sigma_{ij}$ is the differential
scattering cross section $d\sigma_{ij}/d\Omega$ in the centre-of-mass system,
$\theta$ is the scattering angle in that reference system, and
$c \, M_{ij}\,t$, the first argument in $d\sigma_{ij}$ in Eq.(4) of BC,
is to be associated with the total energy $\sqrt{s}$ within the 
center-of-mass system. Furthermore, in comparing results between BC
and GLW one needs to keep in mind that $\sigma_{12}$ quoted in the
diffusion constants given in GLW (computed under the assumption of 
energy- and angle- independent cross sections) is not the total cross
section but that quantity divided by $(4\pi )$.
\end{thebibliography}
\end{document}